%This is Ams-Tex file of a paper
%published in Il Nuovo Cimento B, Vol 111 (11) 1315-1332 (1996)
\input amstex

\catcode`\@=11
\newdimen\@wholewidth
\newdimen\@halfwidth
\newdimen\unitlength \unitlength =1pt
\newbox\@picbox
\newdimen\@picht
\newif\if@ignore
\newdimen\@tempdima
\newdimen\@tempdimb
\newcount\@tempcnta
\newcount\@tempcntb
\newbox\@tempboxa
\newcount\@xarg
\newcount\@yarg
\newcount\@yyarg
\newcount\@multicnt
\newdimen\@xdim
\newdimen\@ydim
\newbox\@linechar
\newdimen\@linelen
\newdimen\@clnwd
\newdimen\@clnht
\newdimen\@dashdim
\newbox\@dashbox
\newcount\@dashcnt

\font\tenln=line10
\font\tenlnw=line10
\font\tencirc=lcircle10
\font\tencircw=lcircle10

\def\thinlines{\let\@linefnt\tenln \let\@circlefnt\tencirc
  \@wholewidth\fontdimen8\tenln \@halfwidth .5\@wholewidth}
\def\thicklines{\let\@linefnt\tenlnw \let\@circlefnt\tencircw
  \@wholewidth\fontdimen8\tenlnw \@halfwidth .5\@wholewidth}

\thinlines

\def\picture(#1,#2){\@ifnextchar({\@picture(#1,#2)}{\@picture(#1,#2)(0,0)}}

\def\@picture(#1,#2)(#3,#4){\@picht #2\unitlength
\setbox\@picbox\hbox to #1\unitlength\bgroup
\hskip -#3\unitlength \lower #4\unitlength \hbox\bgroup\ignorespaces}

\def\endpicture{\egroup\hss\egroup\ht\@picbox\@picht
\dp\@picbox\z@\leavevmode\box\@picbox}

\def\@endparenv{\addpenalty\@endparpenalty\addvspace\@topsepadd\@endpetrue}

\def\@doendpe{\@endpetrue
     \def\par{\@restorepar\everypar{}\par\@endpefalse}\everypar
               {\setbox0=\lastbox\everypar{}\@endpefalse}}

\newif\if@endpe
\@endpefalse

\def\linethickness#1{\@wholewidth #1\relax \@halfwidth .5\@wholewidth}

\def\begin#1{\def\@tempa{\def\@currenvir{#1}%
  \csname #1\endcsname}\begingroup\global\@ignorefalse%
   \let\end=\latexend\let\line=\latexline\@endpefalse\@tempa}

\def\latexend#1{\csname end#1\endcsname%
     \if@endpe\global\let\@gtempa\@doendpe
     \else\global\let\@gtempa\relax\fi
    \endgroup
     \@gtempa
     \if@ignore \global\@ignorefalse
     \ignorespaces\fi}

\def\@ifnextchar#1#2#3{\let\@tempe #1\def\@tempa{#2}\def\@tempb{#3}\futurelet
    \@tempc\@ifnch}
\def\@ifnch{\ifx \@tempc \@sptoken \let\@tempd\@xifnch
      \else \ifx \@tempc \@tempe\let\@tempd\@tempa\else\let\@tempd\@tempb\fi
      \fi \@tempd}

\def\:{\let\@sptoken= } \:  % this makes \@sptoken a space token

\def\:{\@xifnch} \expandafter\def\: {\futurelet\@tempc\@ifnch}

\def\@ifstar#1#2{\@ifnextchar *{\def\@tempa*{#1}\@tempa}{#2}}

\long\def\put(#1,#2)#3{\@killglue\raise#2\unitlength\hbox to \z@{\kern
#1\unitlength #3\hss}\ignorespaces}

\def\@killglue{\unskip\@whiledim \lastskip >\z@\do{\unskip}}

\def\@whilenoop#1{}
\def\@whilenum#1\do #2{\ifnum #1\relax #2\relax\@iwhilenum{#1\relax
     #2\relax}\fi}
\def\@iwhilenum#1{\ifnum #1\let\@nextwhile=\@iwhilenum
         \else\let\@nextwhile=\@whilenoop\fi\@nextwhile{#1}}
\def\@whiledim#1\do #2{\ifdim #1\relax#2\@iwhiledim{#1\relax#2}\fi}
\def\@iwhiledim#1{\ifdim #1\let\@nextwhile=\@iwhiledim
        \else\let\@nextwhile=\@whilenoop\fi\@nextwhile{#1}}

\long\def\makebox(#1,#2)[#3]#4{\vbox to#2\unitlength
   {\let\mb@b\vss \let\mb@l\hss\let\mb@r\hss
    \let\mb@t\vss
    \@tfor\@tempa :=#3\do{\expandafter\let
        \csname mb@\@tempa\endcsname\relax}%
\mb@t\hbox to #1\unitlength{\mb@l #4\mb@r}\mb@b}}

\def\@nnil{\@nil}
\def\@empty{}
\def\@fornoop#1\@@#2#3{}

\def\@tfor#1:=#2\do#3{\xdef\@fortmp{#2}\ifx\@fortmp\@empty \else
    \@tforloop#2\@nil\@nil\@@#1{#3}\fi}
\def\@tforloop#1#2\@@#3#4{\def#3{#1}\ifx #3\@nnil
       \let\@nextwhile=\@fornoop \else
      #4\relax\let\@nextwhile=\@tforloop\fi\@nextwhile#2\@@#3{#4}}

\def\rule{\@ifnextchar[{\@rule}{\@rule[\z@]}}

\def\@height{height}
\def\@depth{depth}
\def\@width{width}

\def\@rule[#1]#2#3{\@tempdima#3\advance\@tempdima #1\leavevmode\hbox{\vrule
  \@width#2 \@height\@tempdima \@depth-#1}}

\long\def\framebox(#1,#2)[#3]#4{\frame{\makebox(#1,#2)[#3]{#4}}}

\long\def\frame#1{\leavevmode
    \hbox{\hskip-\@wholewidth
     \vbox{\vskip-\@wholewidth
            \hrule \@height\@wholewidth
          \hbox{\vrule \@width\@wholewidth #1\vrule \@width\@wholewidth}\hrule
           \@height \@wholewidth\vskip -\@halfwidth}\hskip-\@wholewidth}}

\def\dashbox#1(#2,#3){\leavevmode\hbox to \z@{\baselineskip \z@%
\lineskip \z@%
\@dashdim=#2\unitlength%
\@dashcnt=\@dashdim \advance\@dashcnt 200
\@dashdim=#1\unitlength\divide\@dashcnt \@dashdim
\ifodd\@dashcnt\@dashdim=\z@%
\advance\@dashcnt \@ne \divide\@dashcnt \tw@
\else \divide\@dashdim \tw@ \divide\@dashcnt \tw@
\advance\@dashcnt \m@ne
\setbox\@dashbox=\hbox{\vrule \@height \@halfwidth \@depth \@halfwidth
\@width \@dashdim}\put(0,0){\copy\@dashbox}%
\put(0,#3){\copy\@dashbox}%
\put(#2,0){\hskip-\@dashdim\copy\@dashbox}%
\put(#2,#3){\hskip-\@dashdim\box\@dashbox}%
\multiply\@dashdim 3
\fi
\setbox\@dashbox=\hbox{\vrule \@height \@halfwidth \@depth \@halfwidth
\@width #1\unitlength\hskip #1\unitlength}\@tempcnta=0
\put(0,0){\hskip\@dashdim \@whilenum \@tempcnta <\@dashcnt
\do{\copy\@dashbox\advance\@tempcnta \@ne }}\@tempcnta=0
\put(0,#3){\hskip\@dashdim \@whilenum \@tempcnta <\@dashcnt
\do{\copy\@dashbox\advance\@tempcnta \@ne }}%
\@dashdim=#3\unitlength%
\@dashcnt=\@dashdim \advance\@dashcnt 200
\@dashdim=#1\unitlength\divide\@dashcnt \@dashdim
\ifodd\@dashcnt \@dashdim=\z@%
\advance\@dashcnt \@ne \divide\@dashcnt \tw@
\else
\divide\@dashdim \tw@ \divide\@dashcnt \tw@
\advance\@dashcnt \m@ne
\setbox\@dashbox\hbox{\hskip -\@halfwidth
\vrule \@width \@wholewidth
\@height \@dashdim}\put(0,0){\copy\@dashbox}%
\put(#2,0){\copy\@dashbox}%
\put(0,#3){\lower\@dashdim\copy\@dashbox}%
\put(#2,#3){\lower\@dashdim\copy\@dashbox}%
\multiply\@dashdim 3
\fi
\setbox\@dashbox\hbox{\vrule \@width \@wholewidth
\@height #1\unitlength}\@tempcnta0
\put(0,0){\hskip -\@halfwidth \vbox{\@whilenum \@tempcnta < \@dashcnt
\do{\vskip #1\unitlength\copy\@dashbox\advance\@tempcnta \@ne }%
\vskip\@dashdim}}\@tempcnta0
\put(#2,0){\hskip -\@halfwidth \vbox{\@whilenum \@tempcnta< \@dashcnt
\relax\do{\vskip #1\unitlength\copy\@dashbox\advance\@tempcnta \@ne }%
\vskip\@dashdim}}}\makebox(#2,#3)}

\newif\if@negarg

\def\latexline(#1,#2)#3{\@xarg #1\relax \@yarg #2\relax
\@linelen=#3\unitlength
\ifnum\@xarg =0 \@vline
  \else \ifnum\@yarg =0 \@hline \else \@sline\fi
\fi}

\def\@sline{\ifnum\@xarg< 0 \@negargtrue \@xarg -\@xarg \@yyarg -\@yarg
  \else \@negargfalse \@yyarg \@yarg \fi
\ifnum \@yyarg >0 \@tempcnta\@yyarg \else \@tempcnta -\@yyarg \fi
\ifnum\@tempcnta>6 \@badlinearg\@tempcnta0 \fi
\ifnum\@xarg>6 \@badlinearg\@xarg 1 \fi
\setbox\@linechar\hbox{\@linefnt\@getlinechar(\@xarg,\@yyarg)}%
\ifnum \@yarg >0 \let\@upordown\raise \@clnht\z@
   \else\let\@upordown\lower \@clnht \ht\@linechar\fi
\@clnwd=\wd\@linechar
\if@negarg \hskip -\wd\@linechar \def\@tempa{\hskip -2\wd\@linechar}\else
     \let\@tempa\relax \fi
\@whiledim \@clnwd <\@linelen \do
  {\@upordown\@clnht\copy\@linechar
   \@tempa
   \advance\@clnht \ht\@linechar
   \advance\@clnwd \wd\@linechar}%
\advance\@clnht -\ht\@linechar
\advance\@clnwd -\wd\@linechar
\@tempdima\@linelen\advance\@tempdima -\@clnwd
\@tempdimb\@tempdima\advance\@tempdimb -\wd\@linechar
\if@negarg \hskip -\@tempdimb \else \hskip \@tempdimb \fi
\multiply\@tempdima \@m
\@tempcnta \@tempdima \@tempdima \wd\@linechar \divide\@tempcnta \@tempdima
\@tempdima \ht\@linechar \multiply\@tempdima \@tempcnta
\divide\@tempdima \@m
\advance\@clnht \@tempdima
\ifdim \@linelen <\wd\@linechar
   \hskip \wd\@linechar
  \else\@upordown\@clnht\copy\@linechar\fi}

\def\@hline{\ifnum \@xarg <0 \hskip -\@linelen \fi
\vrule \@height \@halfwidth \@depth \@halfwidth \@width \@linelen
\ifnum \@xarg <0 \hskip -\@linelen \fi}

\def\@vline{\ifnum \@yarg <0 \@downline \else \@upline\fi}

\def\@upline{\hbox to \z@{\hskip -\@halfwidth \vrule \@width \@wholewidth
   \@height \@linelen \@depth \z@\hss}}

\def\@downline{\hbox to \z@{\hskip -\@halfwidth \vrule \@width \@wholewidth
   \@height \z@ \@depth \@linelen \hss}}

\def\@getlinechar(#1,#2){\@tempcnta#1\relax\multiply\@tempcnta 8
\advance\@tempcnta -9 \ifnum #2>0 \advance\@tempcnta #2\relax\else
\advance\@tempcnta -#2\relax\advance\@tempcnta 64 \fi
\char\@tempcnta}

\def\vector(#1,#2)#3{\@xarg #1\relax \@yarg #2\relax
\@tempcnta \ifnum\@xarg<0 -\@xarg\else\@xarg\fi
\ifnum\@tempcnta<5\relax
\@linelen=#3\unitlength
\ifnum\@xarg =0 \@vvector
  \else \ifnum\@yarg =0 \@hvector \else \@svector\fi
\fi
\else\@badlinearg\fi}

\def\@hvector{\@hline\hbox to 0pt{\@linefnt
\ifnum \@xarg <0 \@getlarrow(1,0)\hss\else
    \hss\@getrarrow(1,0)\fi}}

\def\@vvector{\ifnum \@yarg <0 \@downvector \else \@upvector \fi}

\def\@svector{\@sline
\@tempcnta\@yarg \ifnum\@tempcnta <0 \@tempcnta=-\@tempcnta\fi
\ifnum\@tempcnta <5
  \hskip -\wd\@linechar
  \@upordown\@clnht \hbox{\@linefnt  \if@negarg
  \@getlarrow(\@xarg,\@yyarg) \else \@getrarrow(\@xarg,\@yyarg) \fi}%
\else\@badlinearg\fi}

\def\@getlarrow(#1,#2){\ifnum #2 =\z@ \@tempcnta='33\else
\@tempcnta=#1\relax\multiply\@tempcnta \sixt@@n \advance\@tempcnta
-9 \@tempcntb=#2\relax\multiply\@tempcntb \tw@
\ifnum \@tempcntb >0 \advance\@tempcnta \@tempcntb\relax
\else\advance\@tempcnta -\@tempcntb\advance\@tempcnta 64
\fi\fi\char\@tempcnta}

\def\@getrarrow(#1,#2){\@tempcntb=#2\relax
\ifnum\@tempcntb < 0 \@tempcntb=-\@tempcntb\relax\fi
\ifcase \@tempcntb\relax \@tempcnta='55 \or
\ifnum #1<3 \@tempcnta=#1\relax\multiply\@tempcnta
24 \advance\@tempcnta -6 \else \ifnum #1=3 \@tempcnta=49
\else\@tempcnta=58 \fi\fi\or
\ifnum #1<3 \@tempcnta=#1\relax\multiply\@tempcnta
24 \advance\@tempcnta -3 \else \@tempcnta=51\fi\or
\@tempcnta=#1\relax\multiply\@tempcnta
\sixt@@n \advance\@tempcnta -\tw@ \else
\@tempcnta=#1\relax\multiply\@tempcnta
\sixt@@n \advance\@tempcnta 7 \fi\ifnum #2<0 \advance\@tempcnta 64 \fi
\char\@tempcnta}

\def\@upvector{\@upline\setbox\@tempboxa\hbox{\@linefnt\char'66}\raise
     \@linelen \hbox to\z@{\lower \ht\@tempboxa\box\@tempboxa\hss}}

\def\@downvector{\@downline\lower \@linelen
      \hbox to \z@{\@linefnt\char'77\hss}}

\def\@badlinearg{}

\newif\if@ovt
\newif\if@ovb
\newif\if@ovl
\newif\if@ovr
\newdimen\@ovxx
\newdimen\@ovyy
\newdimen\@ovdx
\newdimen\@ovdy
\newdimen\@ovro
\newdimen\@ovri

\def\@getcirc#1{\@tempdima #1\relax \advance\@tempdima 2pt\relax
  \@tempcnta\@tempdima
  \@tempdima 4pt\relax \divide\@tempcnta\@tempdima
  \ifnum \@tempcnta > 10\relax \@tempcnta 10\relax\fi
  \ifnum \@tempcnta >\z@ \advance\@tempcnta\m@ne
    \else \@warning{Oval too small}\fi
  \multiply\@tempcnta 4\relax
  \setbox \@tempboxa \hbox{\@circlefnt
  \char \@tempcnta}\@tempdima \wd \@tempboxa}

\def\@put#1#2#3{\raise #2\hbox to \z@{\hskip #1#3\hss}}

\def\oval(#1,#2){\@ifnextchar[{\@oval(#1,#2)}{\@oval(#1,#2)[]}}

\def\@oval(#1,#2)[#3]{\begingroup\boxmaxdepth \maxdimen
  \@ovttrue \@ovbtrue \@ovltrue \@ovrtrue
  \@tfor\@tempa :=#3\do{\csname @ov\@tempa false\endcsname}\@ovxx
  #1\unitlength \@ovyy #2\unitlength
  \@tempdimb \ifdim \@ovyy >\@ovxx \@ovxx\else \@ovyy \fi
  \advance \@tempdimb -2pt\relax  %%%% added 7 Dec 89
  \@getcirc \@tempdimb
  \@ovro \ht\@tempboxa \@ovri \dp\@tempboxa
  \@ovdx\@ovxx \advance\@ovdx -\@tempdima \divide\@ovdx \tw@
  \@ovdy\@ovyy \advance\@ovdy -\@tempdima \divide\@ovdy \tw@
  \@circlefnt \setbox\@tempboxa
  \hbox{\if@ovr \@ovvert32\kern -\@tempdima \fi
  \if@ovl \kern \@ovxx \@ovvert01\kern -\@tempdima \kern -\@ovxx \fi
  \if@ovt \@ovhorz \kern -\@ovxx \fi
  \if@ovb \raise \@ovyy \@ovhorz \fi}\advance\@ovdx\@ovro
  \advance\@ovdy\@ovro \ht\@tempboxa\z@ \dp\@tempboxa\z@
  \@put{-\@ovdx}{-\@ovdy}{\box\@tempboxa}%
  \endgroup}

\def\@ovvert#1#2{\vbox to \@ovyy{%
    \if@ovb \@tempcntb \@tempcnta \advance \@tempcntb by #1\relax
      \kern -\@ovro \hbox{\char \@tempcntb}\nointerlineskip
    \else \kern \@ovri \kern \@ovdy \fi
    \leaders\vrule width \@wholewidth\vfil \nointerlineskip
    \if@ovt \@tempcntb \@tempcnta \advance \@tempcntb by #2\relax
      \hbox{\char \@tempcntb}%
    \else \kern \@ovdy \kern \@ovro \fi}}

\def\@ovhorz{\hbox to \@ovxx{\kern \@ovro
    \if@ovr \else \kern \@ovdx \fi
    \leaders \hrule height \@wholewidth \hfil
    \if@ovl \else \kern \@ovdx \fi
    \kern \@ovri}}

\def\circle{\@ifstar{\@dot}{\@circle}}
\def\@circle#1{\begingroup \boxmaxdepth \maxdimen \@tempdimb #1\unitlength
   \ifdim \@tempdimb >15.5pt\relax \@getcirc\@tempdimb
      \@ovro\ht\@tempboxa
     \setbox\@tempboxa\hbox{\@circlefnt
      \advance\@tempcnta\tw@ \char \@tempcnta
      \advance\@tempcnta\m@ne \char \@tempcnta \kern -2\@tempdima
      \advance\@tempcnta\tw@
      \raise \@tempdima \hbox{\char\@tempcnta}\raise \@tempdima
        \box\@tempboxa}\ht\@tempboxa\z@ \dp\@tempboxa\z@
      \@put{-\@ovro}{-\@ovro}{\box\@tempboxa}%
   \else  \@circ\@tempdimb{96}\fi\endgroup}

\def\@dot#1{\@tempdimb #1\unitlength \@circ\@tempdimb{112}}

\def\@circ#1#2{\@tempdima #1\relax \advance\@tempdima .5pt\relax
   \@tempcnta\@tempdima \@tempdima 1pt\relax
   \divide\@tempcnta\@tempdima
   \ifnum\@tempcnta > 15\relax \@tempcnta 15\relax \fi
   \ifnum \@tempcnta >\z@ \advance\@tempcnta\m@ne\fi
   \advance\@tempcnta #2\relax
   \@circlefnt \char\@tempcnta}

\def\emline#1#2#3#4#5#6{%
       \put(#1,#2){\special{em:moveto}}%
       \put(#4,#5){\special{em:lineto}}}

\def\newpic#1{}

\def\newcount{\alloc@0\count\countdef\insc@unt}
\def\setcounter#1#2{{\global\csname c@#1\endcsname#2\relax}}

\def\newcounter#1{\@definecounter{#1}}

\def\@definecounter#1{\expandafter\newcount\csname c@#1\endcsname
     \setcounter{#1}0 \expandafter\gdef\csname cl@#1\endcsname{}
     \expandafter\gdef\csname p@#1\endcsname{}\expandafter
     \gdef\csname the#1\endcsname{\arabic{#1}}}

\newcounter{@sc}
\newcounter{@scp}
\newcounter{@t}
\newskip{\@x}
\newskip{\@xa}
\newskip{\@xb}
\newskip{\@y}
\newskip{\@ya}
\newskip{\@yb}
\newbox{\@pt}
\def\bezier#1(#2,#3)(#4,#5)(#6,#7){\c@@sc#1\relax
 \c@@scp\c@@sc \advance\c@@scp\@ne
 \@xb #4\unitlength \advance\@xb -#2\unitlength \multiply\@xb \tw@
 \@xa #6\unitlength \advance\@xa -#2\unitlength
 \advance\@xa -\@xb \divide\@xa\c@@sc
 \@yb #5\unitlength \advance\@yb -#3\unitlength \multiply\@yb \tw@
 \@ya #7\unitlength \advance\@ya -#3\unitlength
 \advance\@ya -\@yb \divide\@ya\c@@sc
 \setbox\@pt\hbox{\vrule height\@halfwidth depth\@halfwidth 
 width\@wholewidth}\c@@t\z@ 
 \put(#2,#3){\@whilenum{\c@@t<\c@@scp}\do
 {\@x\c@@t\@xa \advance\@x\@xb \divide\@x\c@@sc \multiply\@x\c@@t 
 \@y\c@@t\@ya \advance\@y\@yb \divide\@y\c@@sc \multiply\@y\c@@t 
 \raise \@y \hbox to \z@{\hskip \@x\unhcopy\@pt\hss}\advance\c@@t\@ne}}}

\catcode`\@=\active
\documentstyle{amsppt}
{\catcode`\@=11\gdef\logo@{}}
\TagsOnRight
\loadbold
\pageheight{25 true cm}
\pagewidth{17 true cm}
\document
\centerline{\bf GAUGE-POTENTIAL APPROACH TO THE KINEMATICS OF A MOVING CAR}
\vskip 2cm
\centerline{Mari\'an Fecko ${}^{a)}$}
\centerline{Department of Theoretical Physics, Comenius University}
\centerline{Mlynsk\'a dolina F2, 842 15 Bratislava, Slovakia}
\vskip 3cm
{\bf Abstract}
\vskip 1cm
A kinematics of the motion of a car is reformulated in terms of the theory
of gauge potentials. E(2)-gauge structure originates
in the no-slipping contact of the car with a road.
\vskip 4cm
{\bf 1. Introduction}
\vskip 1cm
\indent
The physically most important field where the mathematical theory of
connections ($\leftrightarrow$ gauge potentials $\leftrightarrow$
Yang-Mills potentials) is used with great success is undoubtedly the
theory of elementary particles. Since, however, the concepts involved are
rather abstract and (especially for a newcomer in the field) mixed
with a number of other (equally abstract) ones, one should
appreciate to find out that gauge potentials can be used in
'much more mundane, but in return more readily visualized, context' [1],
too, viz. in the context of classical mechanics [1],[3],[4],[5] or
hydrodynamics [2]. A nice example of this sort is given in [1] (cf.also [3]).
It was shown there that the natural kinematical framework for computing the
net rotation of a (deformable) body due to a sequence of deformations is the
non-Abelian gauge structure over the space of shapes of the body.
\newline
\indent
In this paper we show that (and rather in detail how) the kinematics of a motion
of a car on a road can be reformulated in terms of non-Abelian gauge
potentials, too. The gauge group is $E(2)$, the Euclidean group
of the translations and rotations of the 2-dimensional plane.
\newline
\indent
It should be noted that the differential geometric treatment of the
car's kinematics was given before in [6]. The new point here is the
addition of the degree of freedom $\alpha$ (see Sec.2) which makes it
possible then to treat the problem in the language of connections.
\newline
\indent
Finally let us mention a technical simplification made in computations. As
is well known, the front (as well as rear) wheels of a car do not rotate
with the same angular velocity in general (the device called {\it
differential} is needed). When we speak about the angle $\alpha$ as being
the angle measuring the orientation of the front wheel, the {\it average}
angle is understood in fact. Or, equivalently, we compute everything as if the
car was a {\it tricycle} (then $\alpha$ is the angle of {\it the} front
wheel). The full account of the situation with two wheels can be done, of
course, but it does not bring anything conceptually new.

\newpage

{\bf 2. The configuration space of a car as a principal E(2)-bundle}
\vskip 1cm
Let $P$ be the configuration space of a car. The coordinates $(\alpha,
\beta,x,y,\varphi)$ are introduced according to the Fig.1, Fig.2, with the
following meaning : $(x,y)$ are the Cartesian coordinates of the center of
the front axle, $\varphi$ is the angle between the $x_1$ axis and the tie
rod ('if that is the name of the thing connecting the front and rear axles'
[6]; it measures the direction in which the car is headed), $\alpha$
measures the orientation of the front wheel with respect to the axle and
$\beta$ is the angle made by the front axle with the tie rod. Thus
$(x,y,\varphi)$ carry the information about the position of the tie rod
alone in the $x_1x_2$ - plane irrespective of the 'shape' of the car
whereas $(\alpha,\beta)$ encode the car's shape regardless of the position
of the tie rod in the $x_1x_2$ - plane.
\vskip 1cm
\midinsert
\special{em:linewidth 0.4pt}
\unitlength 1.00mm
\linethickness{0.4pt}
\begin{picture}(60.00,58.47)
\put(60.00,7.08){\vector(1,0){0.2}}
\emline{0.00}{7.08}{1}{60.00}{7.08}{2}
\put(2.00,55.08){\vector(0,1){0.2}}
\emline{2.00}{5.08}{3}{2.00}{55.08}{4}
\put(3.00,54.08){\makebox(0,0)[lb]{$x_2$}}
\put(57.00,9.08){\makebox(0,0)[lc]{$x_1$}}
\emline{6.00}{7.08}{5}{7.00}{8.08}{6}
\emline{8.00}{9.08}{7}{9.00}{10.08}{8}
\emline{10.00}{11.08}{9}{11.00}{12.08}{10}
\emline{12.00}{13.08}{11}{13.00}{14.08}{12}
\emline{14.00}{15.08}{13}{15.00}{16.08}{14}
\emline{16.00}{17.08}{15}{17.00}{18.08}{16}
\put(27.00,28.08){\vector(1,1){0.2}}
\emline{18.00}{19.08}{17}{27.00}{28.08}{18}
\emline{44.00}{45.08}{19}{45.00}{46.08}{20}
\emline{46.00}{47.08}{21}{47.00}{48.08}{22}
\emline{48.00}{49.08}{23}{49.00}{50.08}{24}
\emline{50.00}{51.08}{25}{51.00}{52.08}{26}
\emline{52.00}{53.08}{27}{53.00}{54.08}{28}
\emline{54.00}{55.08}{29}{55.00}{56.08}{30}
\emline{18.00}{19.08}{31}{27.00}{10.08}{32}
\emline{9.00}{28.08}{33}{18.00}{19.08}{34}
\emline{30.00}{49.08}{35}{32.00}{53.08}{36}
\emline{43.00}{44.08}{37}{31.00}{51.08}{38}
\emline{55.00}{37.08}{39}{43.00}{44.08}{40}
\emline{54.00}{35.08}{41}{56.00}{39.08}{42}
\emline{43.49}{45.31}{43}{44.24}{46.44}{44}
\emline{44.74}{47.69}{45}{45.49}{48.82}{46}
\emline{46.00}{50.07}{47}{46.75}{51.20}{48}
\emline{47.25}{52.45}{49}{48.00}{53.58}{50}
\emline{48.50}{54.83}{51}{49.25}{55.96}{52}
\put(49.38,52.95){\makebox(0,0)[cc]{$\beta$}}
\put(16.04,17.11){\vector(-1,1){0.2}}
\bezier{48}(20.05,7.09)(20.05,13.60)(16.04,17.11)
\put(15.16,10.84){\makebox(0,0)[cc]{$\varphi$}}
\put(23.56,32.02){\makebox(0,0)[cc]{$\vec e$}}
\emline{56.00}{57.08}{53}{57.00}{58.08}{54}
\emline{49.75}{57.34}{55}{50.50}{58.47}{56}
\emline{27.00}{28.08}{57}{43.00}{44.08}{58}
\put(49.87,57.21){\vector(-2,1){0.2}}
\bezier{20}(53.51,54.58)(52.38,56.96)(49.87,57.21)
\put(41.60,37.79){\makebox(0,0)[lc]{(x,y)}}
\put(3.00,1.00){\makebox(0,0)[lc]
{Fig.1 : The coordinates $x$, $y$, $\varphi$, $\beta$ .}}
\emline{7.27}{26.32}{59}{10.65}{29.70}{60}
\emline{25.44}{8.40}{61}{28.57}{11.65}{62}
\end{picture}
\hskip 3cm
\special{em:linewidth 0.4pt}
\unitlength 1.00mm
\linethickness{0.4pt}
\begin{picture}(57.02,55.51)
\emline{29.95}{46.87}{1}{31.18}{46.83}{2}
\emline{31.18}{46.83}{3}{32.40}{46.72}{4}
\emline{32.40}{46.72}{5}{33.62}{46.53}{6}
\emline{33.62}{46.53}{7}{34.82}{46.26}{8}
\emline{34.82}{46.26}{9}{36.00}{45.93}{10}
\emline{36.00}{45.93}{11}{37.16}{45.52}{12}
\emline{37.16}{45.52}{13}{38.29}{45.04}{14}
\emline{38.29}{45.04}{15}{39.39}{44.49}{16}
\emline{39.39}{44.49}{17}{40.45}{43.88}{18}
\emline{40.45}{43.88}{19}{41.47}{43.20}{20}
\emline{41.47}{43.20}{21}{42.45}{42.46}{22}
\emline{42.45}{42.46}{23}{43.39}{41.66}{24}
\emline{43.39}{41.66}{25}{44.27}{40.80}{26}
\emline{44.27}{40.80}{27}{45.09}{39.89}{28}
\emline{45.09}{39.89}{29}{45.86}{38.93}{30}
\emline{45.86}{38.93}{31}{46.57}{37.93}{32}
\emline{46.57}{37.93}{33}{47.22}{36.89}{34}
\emline{47.22}{36.89}{35}{47.80}{35.80}{36}
\emline{47.80}{35.80}{37}{48.31}{34.69}{38}
\emline{48.31}{34.69}{39}{48.75}{33.54}{40}
\emline{48.75}{33.54}{41}{49.12}{32.37}{42}
\emline{49.12}{32.37}{43}{49.42}{31.18}{44}
\emline{49.42}{31.18}{45}{49.64}{29.97}{46}
\emline{49.64}{29.97}{47}{49.79}{28.75}{48}
\emline{49.79}{28.75}{49}{49.87}{27.52}{50}
\emline{49.87}{27.52}{51}{49.86}{26.29}{52}
\emline{49.86}{26.29}{53}{49.79}{25.07}{54}
\emline{49.79}{25.07}{55}{49.63}{23.85}{56}
\emline{49.63}{23.85}{57}{49.41}{22.64}{58}
\emline{49.41}{22.64}{59}{49.10}{21.45}{60}
\emline{49.10}{21.45}{61}{48.73}{20.28}{62}
\emline{48.73}{20.28}{63}{48.28}{19.14}{64}
\emline{48.28}{19.14}{65}{47.77}{18.02}{66}
\emline{47.77}{18.02}{67}{47.18}{16.94}{68}
\emline{47.18}{16.94}{69}{46.53}{15.90}{70}
\emline{46.53}{15.90}{71}{45.82}{14.90}{72}
\emline{45.82}{14.90}{73}{45.05}{13.94}{74}
\emline{45.05}{13.94}{75}{44.22}{13.04}{76}
\emline{44.22}{13.04}{77}{43.33}{12.18}{78}
\emline{43.33}{12.18}{79}{42.40}{11.39}{80}
\emline{42.40}{11.39}{81}{41.42}{10.65}{82}
\emline{41.42}{10.65}{83}{40.39}{9.97}{84}
\emline{40.39}{9.97}{85}{39.33}{9.36}{86}
\emline{39.33}{9.36}{87}{38.22}{8.82}{88}
\emline{38.22}{8.82}{89}{37.09}{8.34}{90}
\emline{37.09}{8.34}{91}{35.93}{7.94}{92}
\emline{35.93}{7.94}{93}{34.75}{7.60}{94}
\emline{34.75}{7.60}{95}{33.55}{7.34}{96}
\emline{33.55}{7.34}{97}{32.33}{7.16}{98}
\emline{32.33}{7.16}{99}{31.11}{7.05}{100}
\emline{31.11}{7.05}{101}{29.88}{7.02}{102}
\emline{29.88}{7.02}{103}{28.65}{7.06}{104}
\emline{28.65}{7.06}{105}{27.43}{7.18}{106}
\emline{27.43}{7.18}{107}{26.22}{7.37}{108}
\emline{26.22}{7.37}{109}{25.02}{7.64}{110}
\emline{25.02}{7.64}{111}{23.84}{7.98}{112}
\emline{23.84}{7.98}{113}{22.68}{8.39}{114}
\emline{22.68}{8.39}{115}{21.55}{8.87}{116}
\emline{21.55}{8.87}{117}{20.45}{9.43}{118}
\emline{20.45}{9.43}{119}{19.39}{10.04}{120}
\emline{19.39}{10.04}{121}{18.37}{10.73}{122}
\emline{18.37}{10.73}{123}{17.39}{11.47}{124}
\emline{17.39}{11.47}{125}{16.46}{12.28}{126}
\emline{16.46}{12.28}{127}{15.59}{13.13}{128}
\emline{15.59}{13.13}{129}{14.76}{14.05}{130}
\emline{14.76}{14.05}{131}{14.00}{15.01}{132}
\emline{14.00}{15.01}{133}{13.29}{16.01}{134}
\emline{13.29}{16.01}{135}{12.65}{17.06}{136}
\emline{12.65}{17.06}{137}{12.07}{18.14}{138}
\emline{12.07}{18.14}{139}{11.56}{19.26}{140}
\emline{11.56}{19.26}{141}{11.13}{20.41}{142}
\emline{11.13}{20.41}{143}{10.76}{21.58}{144}
\emline{10.76}{21.58}{145}{10.47}{22.78}{146}
\emline{10.47}{22.78}{147}{10.25}{23.99}{148}
\emline{10.25}{23.99}{149}{10.10}{25.21}{150}
\emline{10.10}{25.21}{151}{10.03}{26.43}{152}
\emline{10.03}{26.43}{153}{10.04}{27.66}{154}
\emline{10.04}{27.66}{155}{10.12}{28.89}{156}
\emline{10.12}{28.89}{157}{10.28}{30.10}{158}
\emline{10.28}{30.10}{159}{10.51}{31.31}{160}
\emline{10.51}{31.31}{161}{10.82}{32.50}{162}
\emline{10.82}{32.50}{163}{11.19}{33.67}{164}
\emline{11.19}{33.67}{165}{11.65}{34.81}{166}
\emline{11.65}{34.81}{167}{12.16}{35.93}{168}
\emline{12.16}{35.93}{169}{12.75}{37.00}{170}
\emline{12.75}{37.00}{171}{13.41}{38.05}{172}
\emline{13.41}{38.05}{173}{14.12}{39.04}{174}
\emline{14.12}{39.04}{175}{14.90}{40.00}{176}
\emline{14.90}{40.00}{177}{15.73}{40.90}{178}
\emline{15.73}{40.90}{179}{16.62}{41.75}{180}
\emline{16.62}{41.75}{181}{17.55}{42.54}{182}
\emline{17.55}{42.54}{183}{18.54}{43.28}{184}
\emline{18.54}{43.28}{185}{19.57}{43.95}{186}
\emline{19.57}{43.95}{187}{20.64}{44.56}{188}
\emline{20.64}{44.56}{189}{21.74}{45.10}{190}
\emline{21.74}{45.10}{191}{22.87}{45.57}{192}
\emline{22.87}{45.57}{193}{24.04}{45.97}{194}
\emline{24.04}{45.97}{195}{25.22}{46.30}{196}
\emline{25.22}{46.30}{197}{26.42}{46.55}{198}
\emline{26.42}{46.55}{199}{27.64}{46.73}{200}
\emline{27.64}{46.73}{201}{29.95}{46.87}{202}
\emline{29.95}{26.94}{203}{29.95}{28.44}{204}
\emline{29.95}{29.95}{205}{29.95}{31.45}{206}
\emline{29.95}{32.95}{207}{29.95}{34.46}{208}
\emline{29.95}{35.96}{209}{29.95}{37.47}{210}
\emline{29.95}{38.97}{211}{29.95}{40.47}{212}
\emline{29.95}{41.98}{213}{29.95}{43.48}{214}
\emline{29.95}{44.98}{215}{29.95}{46.49}{216}
\emline{29.95}{47.99}{217}{29.95}{49.50}{218}
\emline{29.95}{51.00}{219}{29.95}{52.50}{220}
\emline{29.95}{54.01}{221}{29.95}{55.51}{222}
\emline{3.01}{7.02}{223}{57.02}{7.02}{224}
\put(45.61,39.10){\vector(4,3){0.2}}
\emline{29.95}{26.94}{225}{45.61}{39.10}{226}
\put(41.10,35.59){\vector(3,-4){0.2}}
\bezier{56}(29.95,41.60)(37.47,41.10)(41.10,35.59)
\put(32.21,34.83){\makebox(0,0)[cc]{$\alpha$}}
\put(37.97,29.82){\makebox(0,0)[cc]{$R$}}
\put(-2.00,1.00){\makebox(0,0)[lc]
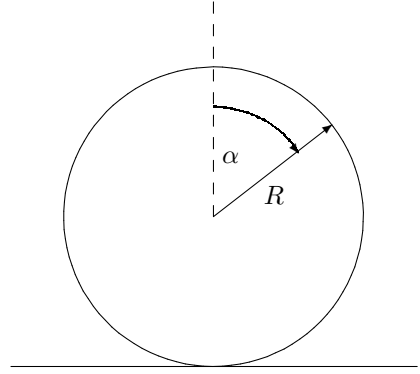
{Fig.2 : The front wheel - the coordinate $\alpha$.}}
\end{picture}

\endinsert

\vskip 1cm
\indent
There is a natural action of the Euclidean group $E(2)$ on $P$, consisting
in 'rigid' motions (rotations and translations) of the car with no change
of its shape, that is to say the motions of the tie rod keeping the shape
fixed. This action $R_{\Cal B} : P\rightarrow P$
(see Appendix A for more technical details) results
in the additional structure of the space $P$, viz. the structure of a
{\it principal fiber bundle} with the group $E(2)$. It is constructed as
follows : two configurations $p,p' \in P$ are declared to be equivalent
if they differ only by a rigid motion from $E(2)$, i.e. if there exists such
$(B,b)\in E(2)$ that the action of $(B,b)$ on $p$ results in $p'$,
i.e. $R_{\Cal B}p=p'$. We
define then $M$ as the factor-space $P/E(2)$, i.e. the points of $M$ are
by definition the equivalence classes in $P$. There is a {\it projection}
map
$$\pi : P \rightarrow M$$
sending the configuration $p$ to its own equivalence class $[p] \equiv
\pi (p) = m$, or in coordinates
$$\pi : (\alpha,\beta,x,y,\varphi) \mapsto (\alpha,\beta)$$
Thus $\pi$ extracts from the complete configuration the information about
the shape of the car and 'forgets' the position of the tie rod within the
$x_1x_2$ - plane.
\newline
\indent
According to the terminology of [1],[2], $P$ is the space of 'located
shapes' whereas $M$ is the space of 'unlocated shapes'.
\newline
\indent
If $m\in M$, the set ${\pi}^{-1}(m) \subset P$ (all those $p\in P$ which
project to the fixed $m\in M$) is called the {\it fiber over} $m$ and here
it represents all configurations ($\equiv$ 'located shapes') sharing the
same ('unlocated') shape. Any two fibers ${\pi}^{-1}(m),{\pi}^{-1}(m')$
are mutually diffeomorphic (equally looking) and their abstract model,
the {\it typical fiber}, is denoted by $\Cal E$ (the space of the locations
of the tie rod) in Appendix A and happens to be diffeomorphic to the group
$E(2)$ itself.

\indent
Notice that the knowledge of the configuration $p\in P$ is equivalent
(globally) to the knowledge of the ordered pair $(m,e)\in M\times \Cal E$.
In other words our {\it total space} $P$ of the bundle is (diffeomorphic
to) the product $M\times \Cal E$ of the {\it base} $M$ and the typical
fiber $\Cal E$
$$P=M\times \Cal E$$
and the bundle projection $\pi$ is realized as a projection
${\pi}_1$ on the first factor
$${\pi}_1 : M\times \Cal E \rightarrow M \hskip 1cm (m,e) \mapsto m$$
This means that our bundle is {\it trivial} (in general this is the case
only locally).
\newline
\indent
The {\it section} of the bundle $\pi : P \rightarrow M$ (the {\it fixation
of the gauge}) is a map
$$\sigma :M\rightarrow P$$
obeying
$$\pi \circ \sigma = \text{identity on} \ M$$
($\sigma(m)$ is to be in the
fiber over $m$). It helps to visualize the abstract
shapes (elements of $M$) localizing each of them somewhere
in the $x_1x_2$ - plane. The convenient (global) section is given in
coordinates by
$$\sigma : (\alpha,\beta) \mapsto (\alpha,\beta,0,0,0) \tag 2.1$$
It realizes all shapes by means of the configurations with the tie rod
situated at the $x_1$-axis to the left with respect to the origin (Fig.3).
Notice
that the coordinates $(x,y,\varphi)$ are closely related (adapted) to this
very section (in fact they are introduced just {\it with respect} to {\it
this} section) :
the section defines (for all $m\in M$) the fiducial
point $\sigma (m)$ in the fiber over $m$. This point is (by
definition) labeled by the coordinates $(\alpha,\beta,0,0,0)$. Then a
general point $p$ in the same fiber (with the same shape) acquires the
coordinates $(\alpha,\beta,
x,y,\varphi)$ if the element $(B,b) \in E(2)$ with
$$B =  \left( \matrix cos \varphi & \sin \varphi \\ - \sin \varphi & cos
\varphi \endmatrix \right) \hskip 1cm b=(x,y)$$
is needed to obtain $p$ from $\sigma (m)$ via the group action.
\vskip 1cm
\hskip 5cm
\special{em:linewidth 0.4pt}
\unitlength 1.00mm
\linethickness{0.4pt}
\begin{picture}(60.00,55.00)
\put(30.00,55.00){\vector(0,1){0.2}}
\emline{30.00}{5.00}{1}{30.00}{55.00}{2}
\put(32.00,53.00){\makebox(0,0)[lb]{$x_2$}}
\put(58.00,26.00){\makebox(0,0)[lc]{$x_1$}}
\emline{3.00}{18.00}{3}{3.00}{42.00}{4}
\put(60.00,30.00){\vector(1,0){0.2}}
\emline{1.00}{30.00}{5}{60.00}{30.00}{6}
\emline{0.00}{42.00}{7}{6.00}{42.00}{8}
\emline{0.00}{18.00}{9}{6.00}{18.00}{10}
\put(16.00,27.00){\makebox(0,0)[cc]{l}}
\emline{25.00}{41.00}{11}{35.00}{19.00}{12}
\emline{25.06}{40.98}{13}{28.20}{42.48}{14}
\emline{21.93}{39.47}{15}{25.06}{40.98}{16}
\emline{34.96}{19.05}{17}{38.10}{20.55}{18}
\emline{31.83}{17.54}{19}{34.96}{19.05}{20}
\emline{29.95}{29.95}{21}{31.95}{30.95}{22}
\emline{33.96}{31.83}{23}{35.96}{32.83}{24}
\emline{37.97}{33.71}{25}{39.97}{34.71}{26}
\emline{41.98}{35.59}{27}{43.98}{36.59}{28}
\emline{45.99}{37.47}{29}{47.99}{38.47}{30}
\emline{50.00}{39.35}{31}{52.01}{40.35}{32}
\put(46.62,37.72){\vector(-3,4){0.2}}
\bezier{36}(49.37,29.95)(49.25,35.09)(46.62,37.72)
\put(44.61,32.83){\makebox(0,0)[cc]{$\beta$}}
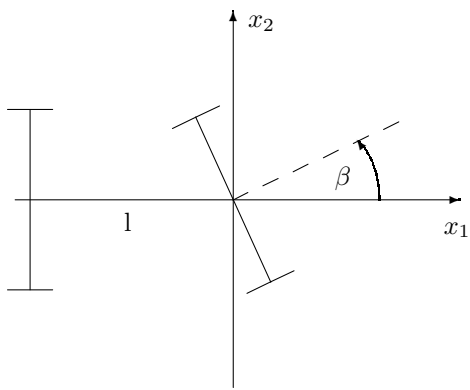
\put(6.00,1.00){\makebox(0,0)[lc]{Fig.3 : The gauge fixation $\sigma$.}}
\end{picture}

\vskip 1cm
The useful possibility is to interpret the section \thetag{2.1} as the
point of view of the {\it driver} (the driver's reference system) : with
respect to his axes ${x'}_1,{x'}_2$ the tie rod is clearly always
at the origin and directed forward ($x \ = y \ = \varphi \ = 0$). Each
other choice of a section (other gauge) corresponds to some different observer,
which can, however, depend on the (unlocated) shape.
\vskip 1cm
\noindent
{\bf 3. The no-slipping contact with a road as a connection on $\pi :P
        \rightarrow M$}
\vskip 1cm
So far we have come to conclusion that the 5-dimensional configuration space $P$
of a car can be treated naturally as a total space of a (trivial) principal
E(2)-bundle $\pi : P \rightarrow M$, $P \equiv M\times \Cal E$. A motion of
the car on a road ($x_1x_2$-plane) is given by a curve
$\gamma (t) \equiv (m (t),e (t))$ on $P \equiv M\times \Cal E$.
 The essential point is, however, that it is only the projection
$m (t) \equiv \pi \circ \gamma (t)$ which is under {\it direct}
control of a driver ($\alpha (t)$ - gas pedal, braces; $\beta (t)$ -
steering wheel). The driver governs directly the 'motion' in the space of
shapes $M$ (the base of the bundle) whereas
what is really his goal is the
change of the position of the tie rod, or in other words to move along
the desired curve $e (t)$ in the typical fiber $\Cal E$ of the bundle.
The necessary 'bridge' between $M$ and $\Cal E$ is given by a system of
(anholonomic) {\it differential constraints} representing physically the
condition of the {\it no-slipping contact} of the wheels with the road.
In such a way the driver's activity represented as the curve
$m (t)$ on $M$ is transformed
to the curve $e (t)$ on $\Cal E$ or, equivalently, $\gamma (t)
\equiv (m (t),e (t))$ on $P$. As we will see, the procedure of
the reconstruction of the complete $\gamma (t)$ on $P$ from its
projection $m
(t)$ on $M$ is just the {\it horizontal lift} $m \mapsto
{m}^h \equiv \gamma$, where the structure necessary for it, viz.
the {\it connection} in the principal bundle $\pi : P\rightarrow M$ (gauge
structure over M) enters the scene as a mathematical expression of the
above mentioned no-slipping contact of the car with the road, i.e. the
constraints of contact can be interpreted in terms of the {\it connection
form} on $P$.
\newline
\indent
In general a connection on a principal fiber bundle $\pi :
P\rightarrow M$ with a group $G$ is given [7] by a $\Cal G$-valued
($\Cal G$ being the Lie algebra of the group $G$) 1-form on $P$, a
connection form. In our
case it means the $3\times 3$ {\it matrix} of 1-forms on $P$ decomposable
with respect to the basis $e_0,e_1,e_2$ of the Lie algebra $e(2)$ of the
group $E(2)$ (see Appendix B)
$$\omega = {\omega}^ae_a = {\omega}^0 e_0 + {\omega}^1e_1 + {\omega}^2e_2=$$
$$= \ \left( \matrix
                          0 & {\omega}^0 & 0 \\
                        - \ {\omega}^0 & 0 & 0 \\
                          {\omega}^1 & {\omega}^2 & 0
                          \endmatrix \right) \tag 3.1$$
where ${\omega}^0,{\omega}^1,{\omega}^2$ are 1-forms on $P$. Thus the
condition of the horizontality
$$\omega =0 \hskip 1cm \text{i.e.} \hskip 1cm
  {\omega}^0={\omega}^1={\omega}^2=0 \tag 3.2$$
represents just 3 independent relations between the differentials $d\alpha,
d\beta,dx,dy,d\varphi$ enabling one to express the infinitesimal changes
$\delta x,\delta y,\delta \varphi$ of the coordinates of the rod in terms of
the given changes $\delta \alpha,\delta \beta$ of the coordinates of the shape
of the car.
\newline
\indent
Note : the equations ${\omega}^a=0$ are not to be interpreted as 1-form
identities on $P$ but rather in the sense that the forms are {\it
annihilated}
(give zero) by the velocity ($\equiv$ tangent) vectors to the real
($\equiv$ obeying the constraints $\Rightarrow$ by definition horizontal)
trajectories on $P$.
\newline
\indent
The computation of the explicit expression for the connection form is
performed in Appendix C. The result reads
$${\omega}^0 = d\varphi - \frac{R}{l}\sin \beta d\alpha$$
$${\omega}^1 = dx+y{\omega}^0-R\cos (\beta +\varphi) d\alpha \tag 3.3$$
$${\omega}^2 = dy-x{\omega}^0-R\sin (\beta +\varphi) d\alpha$$
\newline
\indent
If one fixes the gauge by choosing the section $\sigma$ (Sec.2),
the {\it gauge potential} (in gauge $\sigma$) is given as
$$\Cal A := {\sigma}^* \omega =
  {\Cal A}^ae_a = {\Cal A}^0 e_0 + {\Cal A}^1e_1 + {\Cal A}^2e_2=
  \ \left( \matrix
                          0 & {\Cal A}^0 & 0 \\
                        - \ {\Cal A}^0 & 0 & 0 \\
                          {\Cal A}^1 & {\Cal A}^2 & 0
                          \endmatrix \right) =$$
$$=-\frac Rl
   \left( \matrix
                          0 & \sin \beta & 0 \\
                        - \ \sin \beta & 0 & 0 \\
                     l\cos \beta & l\sin \beta & 0
                          \endmatrix \right)
    d\alpha  \tag 3.4$$
\vskip 1cm
\noindent
{\bf 4. Reconstruction of $\gamma (t)$ on $P$ from $\pi (\gamma
(t))$ on $M$ as a horizontal lift}
\vskip 1cm
The driver's activity is represented by a curve $m (t) \equiv \pi
(\gamma (t))$ on $M$ (a sequence of shapes parametrized by time). The
contact of the wheels with the road results then in a motion in the total
configuration space $P$. According to the meaning of the connection as an
object encoding all constraints of the contact, the resulting trajectory
$\gamma (t)$ on $P$ is the {\it horizontal lift} of the {\it curve} $m(t)$,
i.e. the unique curve ${m}^h(t)$ on $P$ enjoying the following two properties :
\newline
\noindent
$i)$ $\pi ({m}^h (t)) = m (t) \hskip 0.5cm \leftrightarrow \hskip 0.5cm
      {m}^h(t)$ is always exactly 'over' $m (t)$
\newline
\noindent
$ii)$ ${\dot m}^h \ \equiv$ its tangent (velocity) vector  - is
always horizontal, i.e. it annihilates ${\omega}^a$, $a=0,1,2$.
\newline
Let us express these conditions in coordinates. If
$${m}(t) \leftrightarrow (\alpha (t),\beta (t))$$
is given, then its horizontal lift is
$${m}^h(t) \leftrightarrow (\alpha (t),\beta (t),x(t),y(t),\varphi
   (t)).$$
(the same $\alpha$ and $\beta$ are there because of the condition $i)$;
$x,y,\varphi$ are to be determined). Now
$${\dot m}^h(t) = {\dot \alpha}(t){\partial}_{\alpha} +
                       {\dot \beta}(t){\partial}_{\beta} +
                       {\dot x}(t){\partial}_x +
                       {\dot y}(t){\partial}_y +
                       {\dot \varphi}(t){\partial}_{\varphi}$$
and
$$<{\omega}^a,{\dot m}^h(t)> \ = \ 0 \hskip 1cm a=0,1,2$$
gives
$$\dot \varphi = \dot \alpha \frac Rl \sin \beta$$
$$\dot x = \dot \alpha  R \cos (\beta + \varphi) \tag 4.1$$
$$\dot y = \dot \alpha  R \sin (\beta + \varphi)$$
so that
$${\dot m}^h = {\dot \alpha}(t) H_{\alpha} + {\dot \beta}(t) H_{\beta} \tag
  4.2$$
where
$$H_{\alpha} \equiv {\partial}^h_{\alpha} := {\partial}_{\alpha} +
  R\cos(\beta + \varphi) {\partial}_x + R\sin(\beta + \varphi)
  {\partial}_y + \frac Rl \sin \beta {\partial}_{\varphi} \tag 4.3$$
$$H_{\beta} \equiv {\partial}^h_{\beta} := {\partial}_{\beta} \tag 4.4$$
are the {\it horizontal lifts} of the coordinate basis {\it vectors} on
$M$. \newline \indent
The 1-st order linear autonomous system of equations $\thetag{4.1}$, the {\it
parallel transport equations}, solves the reconstruction problem :
given $\alpha (t), \beta (t)$ for $t \in <t_i,t_f>$ (sequence of shapes) and
$(x(t_i),y(t_i),\varphi (t_i)$ (the initial position in the fibre over
$m (t_i)$, i.e. the initial position of the car on the road), it provides
the remaining information about the motion of the car, viz. the sequence of the
positions of the tie rod corresponding to the given sequence of shapes (an
example - the motion with the fixed steering wheel - is computed in
Appendix D). The
parallely transported configuration is then by definition the configuration
${m}^h(t_f)$. (Recall that according to the meaning of the connection
here to follow the parallel transport rule is the same thing as to
be compatible with the constraints of the contact).
\newline
\indent
Note that the equations $\thetag{4.1}$ are invariant (as is the case in general
for the parallel transport equations) with respect to reparametrization -
the speed of the shape sequence is irrelevant, what matters is only the
{\it path} corresponding to $m (t)$ rather then the {\it curve}
$m (t)$ itself. Surprisingly, this rather subtle technical fact seems
to be pretty well known intuitively to our wives when they
prevent us to drive too quickly ('you win nothing by it').
\vskip 1cm
\noindent
{\bf 5. Parking cycles as a clever use of the curvature
        $\Omega$ of the connection}
\vskip 1 cm
To get out of an extremely tight parking spot [6] a
pure translation of the tie rod {\it perpendicular} to the latter, i.e.
(infinitesimally)
$$(x,y,\varphi) \mapsto (x-\epsilon \sin \varphi,y+\epsilon \cos
  \varphi,\varphi) \tag 5.1$$
\noindent
($\epsilon \ll 1$)
is strongly desirable lest we come to contact with
the neighbouring car (and even much worse with the {\it owner} of the
neighbouring car, then).
\newline \indent
On the other hand according to the results of Sec.4. only the motions
generated by some horizontal lift $\dot m^h$ are possible (allowed by the
constraints), i.e. (cf. \thetag{4.2} - \thetag{4.4})
$$(x,y,\varphi) \mapsto (x+\epsilon cos (\varphi +\beta),y+\epsilon \sin
  (\varphi +\beta),\varphi +\frac{\epsilon}{l} \sin \beta)  \tag 5.2$$
\noindent
($\epsilon = \dot \alpha R \delta t \ll 1$).
In no special case \thetag{5.2} reduces to \thetag{5.1} : \thetag{5.2}
consists of both translation and {\it rotation} except for the case
$\beta=0$, when, however, the translation is just {\it along} the tie rod.
Thus it seems that we are simply unlucky and we have to wait until the car
in front of us leaves.
\newline
\indent
This conclusion is, however, too hasty, since we have not used yet the
basic parking algorithm known to every driver, viz. a {\it cycle}
in the space $M$. Let us study for a moment the result of an infinitesimal cycle
of the following structure (see Fig.4) :
\newline \noindent
i) go forth ($\alpha \mapsto \alpha + \epsilon$)
\newline \noindent
ii) turn the steering wheel to the left ($\beta \mapsto \beta + \epsilon$)
\newline \noindent
iii) go (the same step) back ($\alpha + \epsilon \mapsto \alpha$)
\newline \noindent
iv) turn the steering wheel (the same angle) back to the right
    ($\beta + \epsilon \mapsto \beta$)
\newline \noindent
Clearly we finish at (exactly) the same point {\it in} $M$;
the {\it complete} configuration $p\equiv (m,e)$, however, changes :
$p \equiv (m,e) \mapsto (m,e')\equiv p'$, viz. up to the {\it second} order
terms in $\epsilon$ the result is (see below)
$$(\alpha ,\beta ,x,y,\varphi) \mapsto (\alpha ,\beta ,x- {\epsilon}^2R
  \sin (\varphi +\beta),y+{\epsilon}^2R \cos (\varphi +\beta),\varphi +
  {\epsilon}^2\frac{R}{l} \cos \beta)  \tag 5.3$$

\vskip 1cm
\hskip 5cm
\special{em:linewidth 0.4pt}
\unitlength 1.00mm
\linethickness{0.4pt}
\begin{picture}(60.00,55.00)
\put(2.00,55.00){\vector(0,1){0.2}}
\emline{2.00}{5.00}{1}{2.00}{55.00}{2}
\put(60.00,7.00){\vector(1,0){0.2}}
\emline{0.00}{7.00}{3}{60.00}{7.00}{4}
\put(4.00,54.00){\makebox(0,0)[lb]{$\beta$}}
\put(58.00,9.00){\makebox(0,0)[lc]{$\alpha$}}
\put(30.00,20.00){\vector(1,0){0.2}}
\emline{20.00}{20.00}{5}{30.00}{20.00}{6}
\put(40.00,30.00){\vector(0,1){0.2}}
\emline{40.00}{20.00}{7}{40.00}{30.00}{8}
\put(30.00,40.00){\vector(-1,0){0.2}}
\emline{40.00}{40.00}{9}{30.00}{40.00}{10}
\put(20.00,30.00){\vector(0,-1){0.2}}
\emline{20.00}{40.00}{11}{20.00}{30.00}{12}
\emline{20.00}{30.00}{13}{20.00}{20.00}{14}
\emline{30.00}{20.00}{15}{40.00}{20.00}{16}
\emline{40.00}{30.00}{17}{40.00}{40.00}{18}
\emline{30.00}{40.00}{19}{20.00}{40.00}{20}
\put(17.00,17.00){\makebox(0,0)[cc]{($\alpha$,$\beta$)}}
\put(32.00,22.00){\makebox(0,0)[cc]{$\epsilon$}}
\put(42.00,32.00){\makebox(0,0)[cc]{$\epsilon$}}
\put(-2.00,1.00){\makebox(0,0)[lc]{Fig.4 : A simple ifinitesimal parking
    cycle.}}
\end{picture}

\vskip 1cm
Although this does not meet our requirements yet (rotation is present unless
$\beta =\frac{\pi}{2}$; if $\beta$ is $\frac{\pi}{2}$, the translation is
once more {\it along} the tie rod) there is still something interesting here
which will
turn out to be the essential clue for the real solution of the parking
problem. Namely, if one interprets \thetag{5.3} as a {\it direct} step
from $p$ to $p'$ (and not as the {\it effective} one $\equiv$ the result of
the cyclic motion described above), it is {\it forbidden} (it violates the
constraints). This particular step is even forbidden in a 'maximal way' -
it is purely {\it vertical} (projects to the same point in $M$; both $p$
and $p'$ lie in the same fibre). Thus the fact of vital importance is that a
cycle composed exclusively of allowed ($\equiv$ horizontal) steps can result
in the (directly)
forbidden motion ($\equiv$ non-vanishing vertical part). This means that
although we have come
to the conclusion that \thetag{5.1} cannot be realized 'directly' (as one step)
there is still a real hope to produce it effectively - as a result of
(maybe rather involved) cycle of allowed 'simple steps'.
\newline \indent
The most convenient tool for studying the effect of cycles is the language
of vector fields (Appendix E). The cycle i)-iv) above is
just the infinitesimal cycle generated by $H_{\alpha},H_{\beta}$ (on
$P$; its projection to $M$ is generated by ${\partial}_{\alpha},
{\partial}_{\beta}$ and the corresponding loop closes exactly since they
commute); then the
resulting motion \thetag{5.3} follows from the formula (see \thetag{E.1})
$${\chi}^{H_{\beta}}_{- \epsilon} \circ {\chi}^{H_{\alpha}}_{- \epsilon} \circ
  {\chi}^{H_{\beta}}_{\epsilon} \circ {\chi}^{H_{\alpha}}_{\epsilon} =
  {\chi}^{[H_{\alpha},H_{\beta}]}_{- {\epsilon}^2} \tag 5.4$$
and the explicit computation of the needed commutator :
$$[H_{\alpha},H_{\beta}]=R\{ \sin (\varphi +\beta){\partial}_x -
  \cos (\varphi + \beta){\partial}_y -\frac 1l \cos \beta
  {\partial}_{\varphi} \} \tag 5.5$$
Since \thetag{5.3} is not enough, we can try the 'higher' (iterated)
commutators. There are two of them to be computed and the results are :
$$[H_{\alpha},[H_{\alpha},H_{\beta}]] = \frac{R^2}{l}(\cos \varphi
  {\partial}_y - \sin \varphi {\partial}_x ) \tag 5.6$$
$$[H_{\beta},[H_{\alpha},H_{\beta}]] = H_{\alpha} - {\partial}_{\alpha}
  \equiv {\partial}^h_{\alpha} - {\partial}_{\alpha} \tag 5.7$$
Both of these results deserve some attention.
First, note that the r.h.s of \thetag{5.6} just generates the wanted motion
\thetag{5.1} ! In more detail, the identity (see E.3)
$${\chi}^{H_{\beta}}_{- \epsilon} \circ {\chi}^{H_{\alpha}}_{- \epsilon} \circ
  {\chi}^{H_{\beta}}_{ \epsilon} \circ {\chi}^{H_{\alpha}}_{ \epsilon} \circ
  {\chi}^{H_{\alpha}}_{- {\epsilon}^2} \circ {\chi}^{H_{\alpha}}_{- \epsilon} \circ
  {\chi}^{H_{\beta}}_{- \epsilon} \circ {\chi}^{H_{\alpha}}_{ \epsilon} \circ
  {\chi}^{H_{\beta}}_{ \epsilon} \circ {\chi}^{H_{\alpha}}_{{\epsilon}^2} =
  {\chi}^{[{H_{\alpha}},[{H_{\alpha}},{H_{\beta}}]]}_{- {\epsilon}^4} \tag
   5.8$$
tells us that the iterated cycle standing on the l.h.s. of \thetag{5.8}
(try to draw a picture modifying appropriately Fig.4 !) results in
$$(x,y,\varphi) \mapsto (x+{\epsilon}^4 \frac{R^2}{l}\sin
\varphi,y-{\epsilon}^4 \frac{R^2}{l} \cos \varphi,\varphi)  \tag 5.9$$
which is just the pure translation perpendicular to the tie rod. Note that this
type of motion is {\it very} slow and laborious : it is necessary to
perform ten 'simple' steps (of the order $\epsilon$ or ${\epsilon}^2$) to
produce effectively a single step (which is of the order ${\epsilon}^4$)
in the 'right' direction.
\newline
\indent
The same treatment applied to \thetag{5.7} leads to the identity
$${\chi}^{H_{\beta}}_{- \epsilon} \circ {\chi}^{H_{\alpha}}_{- \epsilon} \circ
  {\chi}^{H_{\beta}}_{ \epsilon} \circ {\chi}^{H_{\alpha}}_{ \epsilon} \circ
  {\chi}^{H_{\beta}}_{- {\epsilon}^2} \circ {\chi}^{H_{\alpha}}_{- \epsilon}
  \circ
  {\chi}^{H_{\beta}}_{- \epsilon} \circ {\chi}^{H_{\alpha}}_{ \epsilon} \circ
  {\chi}^{H_{\beta}}_{ \epsilon} \circ {\chi}^{H_{\beta}}_{{\epsilon}^2} =
  {\chi}^{[{H_{\beta}},[{H_{\alpha}},{H_{\beta}}]]}_{- {\epsilon}^4} \tag
  5.10$$
which shows that performing the (iterated) cycle standing on the l.h.s. of
\thetag{5.10} the car moves just like if the driver simply moved forth, but
the front wheel did not rotate (no change of $\alpha$ at all : ideally
slipping contact - ice on the road).
\newline \indent
As mentioned in Appendix E, the possibility of producing 'forbidden'
motions by means of the cycles composed of 'allowed' steps leans heavily
on the fact that the {\it curvature} of the connection in question does not
vanish ($\leftrightarrow$ the horizontal lifts of coordinate basis vectors
do not commute). The (Lie algebra valued) {\it curvature 2-form}, which
happens [7] to be the measure of this non-commutation, can be computed
easily explicitly (using the formula E.8) here and the result reads
$$\Omega \equiv \text{hor} \ d\omega \ =
  {\Omega}^ae_a = {\Omega}^0 e_0 + {\Omega}^1e_1 + {\Omega}^2e_2=$$
$$= \ \left( \matrix
                          0 & {\Omega}^0 & 0 \\
                        - \ {\Omega}^0 & 0 & 0 \\
                          {\Omega}^1 & {\Omega}^2 & 0
                          \endmatrix \right) \tag 5.11$$
where
$${\Omega}^0=\frac Rl \cos \beta d\alpha \wedge d\beta$$
$${\Omega}^1= \{ -R\sin (\beta + \varphi) +\frac Rl y\cos \beta \} d\alpha
\wedge d\beta \tag 5.12$$
$${\Omega}^2= \{R\cos (\beta + \varphi) -\frac Rl x\cos \beta \} d\alpha
\wedge d\beta   $$
If one fixes the gauge by choosing the section $\sigma$ (Sec.2),
the {\it field strength} (in gauge $\sigma$) is given as
$$\Cal F := {\sigma}^* \Omega =
  {\Cal F}^ae_a = {\Cal F}^0 e_0 + {\Cal F}^1e_1 + {\Cal F}^2e_2=
  \ \left( \matrix
                          0 & {\Cal F}^0 & 0 \\
                        - \ {\Cal F}^0 & 0 & 0 \\
                          {\Cal F}^1 & {\Cal F}^2 & 0
                          \endmatrix \right) =$$
$$= \frac Rl
    \left( \matrix
                          0 & \cos \beta & 0 \\
                        - \ \cos \beta & 0 & 0 \\
                    \ \sin \beta & l \ \cos \beta & 0
                          \endmatrix \right)
    d\alpha \wedge d\beta \tag 5.13$$
\vskip 1cm
\noindent
{\bf 6. Particle fields}
\vskip 1cm
The gauge potentials $\Cal A \equiv {\sigma}^*\omega$ (and the field
strengths
$\Cal F \equiv {\sigma}^*\Omega$) do not exhaust all the building blocks
of the gauge theory of elementary particles - there are also
{\it particle fields} there : particles interact via gauge fields (bosons).
\newline \indent
In our model of the kinematics of a car we used only the 'connection part of
the theory' yet. The question arises
whether there is an object here which is described mathematically by a
particle field and whether some standard computation with it does make
sense in this context.
\newline \indent
If $V$ is a vector space in which a representation $\rho$ acts then
[7] particle field of type $\rho$ is a $V$-valued function on $P$
which
transforms according to the representation $\rho$ with respect to the
action of $G$ on $P$; in our model it means
$$\psi : P\rightarrow V \tag 6.1$$
such that
$$\psi (R_{\Cal B}p)= \rho ({\Cal B}^{-1})\psi (p) \tag 6.2$$
Here we give a simple example of such $\psi$. Let $V={\Bbb R}^2$ and define
the function $\psi$ on $P$
$$\psi : (\alpha ,\beta ,x,y,\varphi) \mapsto
  \left( \matrix            \cos \varphi \\
                            \sin \varphi \\
         \endmatrix \right) . \tag 6.3$$
Then $\psi (p)$ just gives the components of the unit vector $\vec e$ fixed
on the car
and directed along the tie rod (Fig.1). According to Appendix A the action
of $E(2)$ has the explicit form
$$(\alpha, \beta,x,y,\varphi) \mapsto R_{\Cal B} (\alpha, \beta, x,y,\varphi)
  \equiv$$
$$\equiv \ (\alpha, \beta, x \cos \Theta - y \sin \Theta
  + b_1, x \sin \Theta + y \cos \Theta + b_2, \varphi +
  \Theta) \tag 6.4$$
and thus if
$$\psi (p) \leftrightarrow  \left( \matrix
                            \cos \varphi \\
                            \sin \varphi \\
                            \endmatrix \right) \tag 6.5$$
then
$$\psi (R_{\Cal B}p) \leftrightarrow  \left( \matrix
                                      \cos (\varphi +\Theta ) \\
                                      \sin (\varphi +\Theta )\\
                                      \endmatrix \right) \ =$$
$$= \           \left( \matrix
                \cos \Theta & -\sin \Theta \\
                \sin \Theta & \cos \Theta \\
                \endmatrix \right)
                \left( \matrix
                \cos \varphi  \\
                \sin \varphi  \\
                \endmatrix \right)  =: \rho ({\Cal B}^{-1})
                \left( \matrix
                \cos \varphi  \\
                \sin \varphi  \\
                \endmatrix \right) \tag 6.6$$
where the representation $\rho$ of $E(2)$ in $V\equiv {\Bbb R}^2$ is given
by
$$\rho (\Cal B) \equiv \rho ( (B(\Theta),b)) =
                \left( \matrix
                \cos \Theta & \sin \Theta \\
               -\sin \Theta & \cos \Theta \\
                \endmatrix \right) \tag 6.7$$
Thus our $\psi$ is the particle field of type $\rho$ given by
\thetag{6.7}.
\newline \indent
Let the motion in $M$ be given by $m(t) \leftrightarrow (\alpha (t),\beta
(t))$. Then the change $\delta \vec e$ of the vector $\vec e$ between $t$
and $t+\delta t$ can be computed as
$$\delta \vec e \leftrightarrow \delta t {\dot m}^h \psi \ = \ \frac{\delta t
\dot \alpha R}{l} \sin \beta
\left( \matrix           - \sin \varphi \\
                            \cos \varphi \\
         \endmatrix \right) \tag 6.8$$
Since $\left( \matrix           - \sin \varphi \\
                            \cos \varphi \\
         \endmatrix \right)$
is just the unit vector orthogonal to $\vec e$, the net angle of rotation
of $\vec e$ is
$$\frac{\delta t \dot \alpha R}{l} \sin \beta \equiv
  \frac{\delta \alpha R}{l} \sin \beta \tag 6.9$$
which can be checked by inspection of Fig.1.
\indent
The same angle can be computed within the gauge fixation $\sigma$, too,
making use of the {\it covariant derivative} of
$$\Phi := {\sigma}^* \psi =
       \left( \matrix  1 \\
                       0 \\
              \endmatrix \right) \leftrightarrow \vec {e_1}$$
viz. (some concepts not mentioned in this paper [7] are needed for it)
$$\delta \Phi = \delta t {\nabla}_{\dot m} ({\sigma}^* \psi ) =
  \delta t <{\sigma}^* D\psi ,\dot m> =$$
$$= \frac{\delta \alpha R}{l} \sin \beta
    \left( \matrix  0 \\
                    1 \\
              \endmatrix \right) \leftrightarrow
              \frac{\delta \alpha R}{l} \sin \beta \vec {e_2}$$
or
$$\delta {\vec {e_1}} = \frac{\delta \alpha R}{l} \sin \beta \vec {e_2}$$
in concord with \thetag{6.9}.
\vskip 1cm
\noindent
{\bf 7. Conclusions and comments}
\vskip 1cm
\indent
In this paper we have presented in some detail a
gauge-theoretic approach to the kinematics of a motion of a car. It
can serve as still another example of application of the ideas and
techniques of the mathematics of gauge fields and related structures within
rather mundane context of the elementary (classical) mechanics (as opposed
to their standard occurrence in 'noble' = 'fundamental' physics).
\newline
\indent
The formal scheme is the same here like in [1] or [3] : there is a 'total'
configuration space ($P$ here $\leftrightarrow$ the space of located shapes
in [1] $\leftrightarrow$ $X$ in [3]) which happens to carry the structure
of the the total space of the principal fibre bundle. The group $G$ acts
there ($E(2)$ here $\leftrightarrow$ $SO(3)$ in [1] $\leftrightarrow$
$SO(d)$ in [3]) and the space of orbits of this action ($M$ here
$\leftrightarrow$ the space of unlocated shapes in [1] $\leftrightarrow$
$\tilde X$ in [3]), the base of the bundle, represents the 'directly
controllable part' of the total configuration space. The connection in
$\pi : P \rightarrow M$ provides the bridge linking the motions in these
two spaces.
\newline
\indent
The main difference lies in the {\it physical origin} of the connection
in question : here (and also in [5]) it encodes the constraints expressing
the no-slipping (direct) {\it contact} of the car with the road whereas in
[1],[3] and [4] it results
from the conservation laws (of the linear as well as the angular momentum)
in '{\it nothing to push against}' situation, i.e. the constraints enter
the problem dynamically.
\vskip 1cm
\noindent
{\bf Appendix A : The action of E(2) on $\Cal E$ and on $P$}
\vskip 1cm
Let $B\in SO(2)$, $b\equiv (b_1,b_2)\in {\Bbb R}^2$. Then one can define
the transformation of the points $\chi \equiv (x_1,x_2) \in {\Bbb R}^2$ by
the couple $(B,b)$ by
$$\chi \mapsto \chi B+b =: {\Cal R}_{(B,b)}\chi \tag A1$$
Geometrically it represents the rotation by $\Theta$ around the origin (if
$B = \left( \matrix \cos \Theta & \sin \Theta \\ - \sin \Theta & \cos \Theta \endmatrix
 \right)$)
followed by the translation by $(b_1,b_2)$, i.e. the {\it Euclidean}
transformation of $\chi$ by $(B,b)\in E(2)$. The rule \thetag{A1} can be
written in purely matrix form (which is advantageous for manipulations with
the gauge potentials) using the following standard trick : let us associate the
$3\times 3$ matrix $\Cal B$ and the row vector $\eta$ with the couple
$(B,b)$ and the row vector $\chi$ respectively according to
$$\Cal B := \left( \matrix B & 0 \\  b & 1 \endmatrix \right) \hskip 1cm
\eta := (\chi,1) \equiv (x_1,x_2,1) \tag A2$$
Then the matrix multiplication of $\eta$ by $\Cal B$ gives
$$\eta \Cal B = {\eta}' = (\chi B + b,1) \equiv ({\Cal R}_{(B,b)}\chi,1) \tag A3$$
i.e. the rule \thetag{A1} is reproduced from the matrix multiplication
of the auxiliary quantities $\eta$ and $\Cal B$.
\newline \indent
The action given by \thetag{A1} or \thetag{A3} transforms the
$(x_1,x_2)$-plane 'rigidly', i.e. all distances are preserved
(${\Cal R}_{\Cal B} \equiv {\Cal R}_{(B,b)}$ is an isometry).
It enables then to define the action ${\hat R}_{\Cal B}$
of E(2) on the space $\Cal E$ of the locations of the tie rod, transforming
simply both
endpoints by ${\Cal R}_{\Cal B}$. If the coordinates $(x,y,\varphi)$ are
introduced to $\Cal E$ according to the Fig.1, one obtains
$$(x,y,\varphi) \mapsto (x \cos \Theta - y \sin \Theta + b_1, x \sin \Theta +
  y \cos \Theta + b_2, \varphi + \Theta) \equiv {\hat R}_{\Cal B} (x,y,\varphi)
  \tag A.4$$
Notice that the general
position $(x,y,\varphi)$ of the rod can be reached from the
reference one $(0,0,0)$ (the rod being situated on the $x_1$-axis left to
the origin) by means of the unique ${\hat R}_{\Cal B}$ :
$${\hat R}_{\Cal B} (0,0,0) = (x,y,\varphi) \hskip 0.5cm \text{for} \hskip
0.5cm  B = \left( \matrix \cos \varphi & \sin \varphi \\ - \sin \varphi & \cos
\varphi \endmatrix \right) \ , \ b=(x,y) \tag A5$$
This means that the action ${\hat R}_{\Cal B}$ is transitive and free and
thus $\Cal E$ is the 'principal E(2)-space'. Note that \thetag{A5} gives the
diffeomorphism of $\Cal E$ and the group E(2) itself , too.
\newline \indent
Finally the action $R_{\Cal B}$ on $P = M \times \Cal E$ is given by
$$(m,e) \mapsto (m,{\hat R}_{\Cal B} e) =: R_{\Cal B}(m,e)$$
or in coordinates
$$(\alpha, \beta,x,y,\varphi) \mapsto R_{\Cal B} (\alpha, \beta, x,y,\varphi)
  \equiv$$
$$\equiv \ (\alpha, \beta, x \cos \Theta - y \sin \Theta
  + b_1, x \sin \Theta + y \cos \Theta + b_2, \varphi +
  \Theta) \tag A.6$$
\vskip 1cm
\noindent
{\bf Appendix B : The Lie algebra $e(2)$ of the group $E(2)$}
\vskip 1cm
According to Appendix A the group $E(2)$ can be realized by the matrices
$\Cal B = \left( \matrix B & 0 \\  b & 1 \endmatrix \right)$, where
$B\in SO(2)$. By definition, the Lie algebra $e(2)$ consists then of all
$3\times 3$ matrices $\Cal C$ such that $1 +\epsilon \ \Cal C \equiv \Cal B
(\epsilon ) \in E(2)$ when the 2-nd order terms in $\epsilon $ are
neglected. This leads to $\Cal C = \left( \matrix C & 0 \\  c & 0 \endmatrix
                          \right)$
with the additional restriction (comming from $B^T B =1$) $C^T = - \ C$,
or explicitly
$$\Cal C = \left( \matrix 0 & {\lambda}_0 & 0 \\
            - {\lambda}_0 & 0 & 0 \\
            {\lambda}_1 & {\lambda}_2 & 0 \endmatrix \right) \hskip 1cm
            {\lambda}_0,{\lambda}_1,{\lambda}_2 \in \Bbb R$$
The matrices
$$e_0 = \left( \matrix 0 & 1 & 0 \\
                      - 1 & 0 & 0 \\
                        0 & 0 & 0 \endmatrix \right) \hskip 1cm
  e_1 = \left( \matrix 0 & 0 & 0 \\
                        0 & 0 & 0 \\
                        1 & 0 & 0 \endmatrix \right) \hskip 1cm
  e_2 = \left( \matrix 0 & 0 & 0 \\
                        0 & 0 & 0 \\
                        0 & 1 & 0 \endmatrix \right) $$
can serve then as the basis of $e(2)$ ; their commutation relations read
$$[e_0,e_1] \ = \ - \ e_2$$
$$[e_0,e_2] \ = \ + \ e_1 \tag B.1$$
$$[e_1,e_2] \ = \ 0$$
and so the only non-zero structure constants are
$$c^2_{10}=-c^2_{01}= c^1_{02}=-c^1_{20}= 1 \tag B.2$$
\vskip 1cm
\noindent
{\bf Appendix C : A computation of the connection form $\omega$}
\vskip 1cm
In general a connection form can be written as follows
$$\omega \ = \ {\Cal B}^{-1} \bar \omega \Cal B \ + \ {\Cal B}^{-1}
  d \Cal B \tag C1$$
where $\bar \omega \equiv {\bar \omega}^a e_a$ is some (yet unknown)
$e(2)$-valued 1-form {\it on $M$} and
$$\Cal B = \left( \matrix \cos \varphi & \sin\varphi & 0 \\
                      - \ \sin \varphi & \cos\varphi & 0 \\
                                           x & y & 1 \endmatrix \right)
           \hskip 0.5cm \in E(2) \tag C2$$
The form $\omega$ defines the horizontal directions (the relations between
$d\alpha ,d\beta ,dx,dy, \text{and} \ d\varphi$ as a result of the
no-slipping contact of the wheels with the road) by the equations
${\omega}^a \ = \ 0$, $a=0,1,2$. In particular at the points of the section
$\sigma (M) \subset P$, corresponding to the 'standard' position of a car
(i.e. for $x=y=\varphi =0$; cf.Sec.2) one has $\Cal B = 1 = {\Cal B}^{-1}$
and
$${\omega}_{\Cal B = 1} \ = \ \bar \omega + {(d\Cal B)}_{\Cal B = 1} \tag C3$$
Thus
$${\omega}_{\Cal B =1} =  \left( \matrix
                          0 & {\bar \omega}^0 & 0 \\
                        - \ {\bar \omega}^0 & 0 & 0 \\
                          {\bar \omega}^1 & {\bar \omega}^2 & 0
                          \endmatrix \right) \
                          + \
                          \left( \matrix
                          0 & d\varphi & 0 \\
                        - \ d\varphi & 0 & 0 \\
                          dx & dy & 0
                          \endmatrix \right) \ =$$
$$= (d\varphi + {\bar \omega}^0)e_0 + (dx + {\bar \omega}^1)e_1 +
   (dy + {\bar \omega}^2)e_2 \tag C4$$
The equation ${\omega}_{\Cal B =1}=0$ by definition singles out
the horizontal directions for $x=y=\varphi =0$; it reads
$$d\varphi = - {\bar \omega}^0$$
$$dx = - {\bar \omega}^1 \tag C5$$
$$dy = - {\bar \omega}^2$$
On the other hand the no-slipping contact constraints for the standard
position $x=y=\varphi =0$ can be easily read out from the Fig.3 :
if $\alpha \mapsto \alpha + \delta \alpha$ ($\delta \alpha \ll 1$), then
$(x,y)\equiv (0,0) \mapsto (\delta \alpha R\cos \beta , \delta \alpha
R\sin \beta) \equiv
(\delta x,\delta y)$, $(-l,0) \mapsto (-l+\delta \alpha R\cos\beta,
0) \Rightarrow \varphi \mapsto \varphi + \delta \alpha \frac{R}{l}
\sin \beta$; if $\beta \mapsto \beta + \delta \beta$ ($\delta
\beta \ll 1$), then $(x,y,\varphi) \mapsto (x,y,\varphi)$. Thus
$$d\varphi = \frac{R}{l}\sin\beta d\alpha$$
$$dx       = R\cos\beta d\alpha \tag C6$$
$$dy       = R\sin\beta d\alpha$$
A comparison with \thetag{C5} gives
$${\bar \omega}^0 = - \frac{R}{l} \sin \beta d \alpha$$
$${\bar \omega}^1 = - R \cos \beta d \alpha \tag C7$$
$${\bar \omega}^2 = - R \sin \beta d \alpha$$
$$\bar \omega \equiv {\bar \omega}^a e_a = -\frac{R}{l}
               \left( \matrix
                          0 & \sin \beta & 0 \\
                        - \sin \beta & 0 & 0 \\
                          l\cos \beta & l\sin \beta & 0
                          \endmatrix \right)
                          d\alpha \tag C8$$
Inserting this into \thetag{C1} leads finally to
$$\omega = {\omega}^0e_0+{\omega}^1e_1+{\omega}^2e_2$$
where
$${\omega}^0 = d\varphi - \frac{R}{l}\sin \beta d\alpha$$
$${\omega}^1 = dx+y{\omega}^0-R\cos (\beta +\varphi) d\alpha \tag C9$$
$${\omega}^2 = dy-x{\omega}^0-R\sin (\beta +\varphi) d\alpha$$
Thus the differential constraints in {\it general} configuration are
(${\omega}^a=0$)
$$d\varphi =  \frac{R}{l}\sin \beta d\alpha$$
$$dx=R\cos (\beta +\varphi) d\alpha \tag C10$$
$$dy=R\sin (\beta +\varphi) d\alpha$$
Note the absence of the differential $d\beta$ on the r.h.s. - it reflects
the evident fact that turning the steering wheel alone results in no motion
of the tie rod.
\vskip 1cm
\noindent
{\bf Appendix D : Motion of the car with fixed steering wheel}
\vskip 1cm
In the case of a fixed steering wheel ($\beta (t) = {\beta}_0 =
\text{const}$) the parallel transport equations $\thetag{4.1}$ read
$${\varphi}'(\alpha)=\frac Rl \sin {\beta}_0$$
$$x'(\alpha)=R\cos ({\beta}_0+\varphi (\alpha)) \tag D.1$$
$$y'(\alpha)=R\sin ({\beta}_0+\varphi (\alpha))$$
\noindent
(${\varphi}'(\alpha) \equiv \frac{d\varphi}{d\alpha},\dots)$. They are
easily solved. If ${\beta}_0 \ne 0$, then
$$\varphi (\alpha) = {\varphi}_0+\alpha \frac Rl \sin {\beta}_0$$
$$x(\alpha)=x_0+\frac{l}{\sin {\beta}_0}
  (\sin (\varphi (\alpha) +{\beta}_0) -
  \sin({\varphi}_0+{\beta}_0))$$
$$y(\alpha)=y_0-\frac{l}{\sin {\beta}_0}
  (\cos (\varphi (\alpha) +{\beta}_0) -
  \cos({\varphi}_0+{\beta}_0))$$
and consequently
$${(x(\alpha)-x_c)}^2+{(y(\alpha)-y_c)}^2=r^2_c$$
where
$$r_c\equiv \frac{l}{\sin {\beta}_0}$$
$$x_c\equiv x_0-r_c\sin ({\varphi}_0+{\beta}_0)$$
$$y_c\equiv x_0+r_c\cos ({\varphi}_0+{\beta}_0)$$
Thus, as expected, the front wheel draws a {\it circle} with the radius
$r_c$ and the center $(x_c,y_c)$.
\newline
If ${\beta}_0=0$, the equations $\thetag{D.1}$ give
$$\varphi (\alpha) = {\varphi}_0$$
$$x(\alpha)=x_0+\alpha R \cos {\varphi}_0$$
$$y(\alpha)=y_0+\alpha R \sin {\varphi}_0$$
which is a {\it straight line} in the direction of the tie rod.
\vskip 1cm
\noindent
{\bf Appendix E : Commutators, infinitesimal cycles and the curvature}
\vskip 1cm
Let $U,V$ be two vector fields on a manifold $\Cal M$, $[U,V]$ their
commutator (Lie bracket) and ${\chi}^U_t$, ${\chi}^V_t$ and ${\chi}^{[U,V]}_t$
the corresponding flows (${\chi}^U_t$ is the map $\Cal M \rightarrow \Cal M$
sending each point $x\in \Cal M$ a (parameter) distance $t$ along the
integral curve of $U$; it holds ${\chi}^U_{(t+s)} = {\chi}^U_t \circ
{\chi}^U_s = {\chi}^U_s \circ {\chi}^U_t$). Then a computation shows that up
to the {\it second order terms} in $\epsilon \ll 1$ the following
important identity is valid :
$${\chi}^V_{- \epsilon} \circ {\chi}^U_{- \epsilon} \circ
  {\chi}^V_{\epsilon} \circ {\chi}^U_{ \epsilon} =
  {\chi}^{[U,V]}_{- {\epsilon}^2} \tag E.1$$
or equivalently
$${\chi}^{[U,V]}_{{\epsilon}^2} \circ {\chi}^V_{- \epsilon} \circ
  {\chi}^U_{- \epsilon} \circ {\chi}^V_{\epsilon} \circ {\chi}^U_{ \epsilon} =
  \text{identity on} \ \Cal M \tag E.2$$
From these formulae one deduces the standard interpretation of the
commutator of two vector fields : the infinitesimal cycle generated by $U$
and $V$ (l.h.s. of \thetag{E.1}) does not end at the original point within
the accuracy ${\epsilon}^2$ (although it {\it does} within the accuracy
$\epsilon$) but rather one has to add one order smaller step along $[U,V]$
to close the loop (l.h.s. of \thetag{E.2}).
\newline \indent
Now if $V$ is itself a commutator, $V=[W,Z]$, the twofold use of
\thetag{E.1} yields
$${\chi}^Z_{- \epsilon} \circ {\chi}^W_{- \epsilon} \circ
  {\chi}^Z_{\epsilon} \circ {\chi}^W_{ \epsilon} \circ
  {\chi}^U_{- {\epsilon}^2} \circ {\chi}^W_{- \epsilon} \circ
  {\chi}^Z_{- \epsilon} \circ {\chi}^W_{ \epsilon} \circ
  {\chi}^Z_{\epsilon} \circ {\chi}^U_{{\epsilon}^2} =
  {\chi}^{[U,[W,Z]]}_{- {\epsilon}^4} \tag E.3$$
Thus the computation of 'simple' ($[U,V]$) and iterated ($[U,[W,Z]]$)
commutators tells us what is the result of a simple (4 steps) and
iterated (4 steps, but two of them being themselves results of 4 steps,
i.e. together 10 simple steps) cycles respectively (the higher iterated
commutators can be treated in the same way).
\newline \indent
All said until now is valid for any vector fields on any manifold. In the
case when the vector fields in question are horizontal lifts, the resulting
commutator can be expressed in terms of the {\it curvature} of the
connection. For doing this we need first the concept of the {\it
fundamental fields} of the action $R_{\Cal B}$. By definition the field
${\xi}_{\Cal C}$, $\Cal C \equiv {\Cal C}^ae_a \in e(2)$, generates the
motion of any point $p$ under the action of the one-parameter subgroup
${\Cal B}(\lambda) = e^{\lambda \Cal C}$, i.e. for $p(\lambda) :=
R_{\Cal B (\lambda)}p$
$${\xi}_{\Cal C}(p) := \dot p (0) \tag E.4$$
For the basis elements $e_0,e_1,e_2 \in e(2)$ we obtain explicitly
$${\xi}_0\equiv {\xi}_{e_0} = -y{\partial}_x + x{\partial}_y
+{\partial}_{\varphi}$$
$${\xi}_1\equiv {\xi}_{e_1} = {\partial}_x \tag E.5$$
$${\xi}_2\equiv {\xi}_{e_2} = {\partial}_y $$
and in general
$${\xi}_{\Cal C} \equiv {\xi}_{{\Cal C}^ae_a} = {\Cal C}^a {\xi}_{e_a}
  \equiv  {\Cal C}^a {\xi}_a \tag E.6$$
These fields are purely {\it vertical} (directed along the fiber), since (by
definition) the action is vertical ($p$ and $R_{\Cal B}p$ lie in the same
fiber for all $p$, ${\Cal B}$). Now the relevant formula for the commutator
of $H_{\alpha}$ and $H_{\beta}$ is
$$[H_{\alpha},H_{\beta}]={\xi}_{-\Omega (H_{\alpha},H_{\beta})} = \
  -{\Omega}^a (H_{\alpha},H_{\beta}) {\xi}_a \tag E.7$$
where the ($e(2)$-valued ) {\it curvature 2-form} $\Omega = {\Omega}^ae_a$
is given by
$${\Omega}^a=d{\omega}^a + \frac 12 c^a_{bc} \ {\omega}^b \wedge {\omega}^c
  \tag E.8$$
($c^a_{bc}$ being the structure constants; they are computed in Appendix
B). The formula \thetag{E.7} shows that
\newline
i) $[H_{\alpha},H_{\beta}]$ is non-zero if and only if $\Omega$ is non-zero
\newline
ii) $[H_{\alpha},H_{\beta}]$ is purely vertical $\Rightarrow$ the
corresponding cycle generates 'forbidden' motion.
\newline
The explicit form of $\Omega$ in our case is displayed in Sec.5 (see
\thetag{5.11} , \thetag{5.12}).
\vskip 1cm
\noindent
{\bf 8. References}
\vskip 1cm
\noindent
$a)$ Present address : Department of Theoretical Physics, Co\-me\-nius
     University, Mlynsk\'a dolina F2, 842 15 Bratislava, Slovakia; e-mail:
     fecko\@fmph.uniba.sk (published in Il Nuovo Cimento B, Vol 111 (11)
     1315-1332 (1996))
\newline
\noindent
[1] A.Shapere, F.Wilczek: "Gauge kinematics of deformable bodies",
    Am.J.Phys.57 (6),514-518, June 1989
\newline
\noindent
[2] A.Shapere, F.Wilczek: "Geometry of Self-Propulsion at Low Reynolds
    number, J.Fluid.Mech. 198, 557-585 (1989)
\newline
\noindent
[3] A.Guichardet: "On rotation and vibration motions of molecules",
    Ann.Inst.Henri Poincar', Vol.40, n.3, 1984, p.329-342
\newline
\noindent
[4] M.Fecko : "Falling cat" connections and the momentum map,
  J.Math.Phys. 36 (12) 6709-6719 (1995) (available also as
  physics/9702010 at http://xxx.lanl.gov )
\newline
\noindent
[5] M.Fecko : U(1)-gauge structure associated with a motion of a guitar
    string, Acta Physica Slovaca vol.44, No.6, 445-449 (1994)
\newline
\noindent
[6] E.Nelson : Tensor analysis, Princeton Univ. Press 1967, p.33-36
\newline
\noindent
[7] A.Trautman : Differential geometry for physicists, Bibliopolis, Napoli,
    1984, p.88-89,102-103
\enddocument
\end